\documentclass[aps,prb,superscriptaddress,showkeys,longbibliography]{revtex4-2}
\usepackage[utf8]{inputenc}
\usepackage{amsmath,amssymb,amsfonts}
\usepackage{multirow}
\usepackage{graphicx,float}
\usepackage[colorlinks=true,
citecolor=blue,
linkcolor=blue,
urlcolor=blue]{hyperref}
\usepackage{footmisc}

\begin{document}

\title{Ab-initio Study of Electronic and Lattice Dynamical Properties of monolayer ZnO under Strain}
\author{Saumen Chaudhuri}
\affiliation{Department of Physics, Indian Institute of Technology Kharagpur, Kharagpur 721302, India}
\author{{A. K. Das}}
\affiliation{Department of Physics, Indian Institute of Technology Kharagpur, Kharagpur 721302, India}
\author{{G. P. Das}}
\email[corresponding author: ]{gourpdas@gmail.com}
\affiliation{Research Institute for Sustainable Energy, TCG Centres for Research and Education in Science and Technology, Sector V, Salt Lake, Kolkata 700091, India}
\affiliation{Department of Physics, St. Xavier's College, 30 Mother Teresa Sarani, Kolkata 700016, India}
\author{B. N. Dev}
\email[corresponding author: ]{bhupen.dev@gmail.com}
\affiliation{Department of Physics and School of Nano Science and Technology, Indian Institute of Technology Kharagpur, Kharagpur, 721302, India}
\affiliation{Centre for Quantum Engineering, Research and Education, TCG Centres for Research and Education in Science and Technology, Sector V, Salt Lake, Kolkata 700091, India}

\begin{abstract}
First-principles density functional theory based calculations have been performed to investigate the strain-induced modifications in the electronic and vibrational properties of monolayer (ML) ZnO. Wide range of in-plane tensile and compressive strains along different directions are applied to analyse the modifications in detail. The electronic band gap reduces under both tensile and compressive strains and a direct to indirect band gap transition occurs for high values of biaxial tensile strain. The relatively low rate of decrease of band gap and large required strain for direct to indirect band gap transition compared to other $2$D materials are analysed. Systematic decrease in the frequency of the in-plane and increase in the out-of-plane optical phonon modes with increasing tensile strain are observed. The in-plane acoustic modes show linear dispersion for unstrained as well as strained cases. However, the out-of-plane acoustic mode (ZA), which shows quadratic dispersion in the unstrained condition, turns linear with strain. The dispersion of the ZA mode is analysed using the shell elasticity theory and the possibility of ripple formation with strain is analysed. The strain-induced linearity of the ZA mode indicates the absence of rippling under strain. Finally, the stability limit of ML-ZnO is investigated and found that for $18\%$ biaxial tensile strain the structure shows instability with the emergence of imaginary phonon modes. Furthermore, the potential of ML-ZnO to be a good thermoelectric material  is analyzed in an intuitive way based on the calculated electronic and phononic properties. Our results, thus, not only highlight the significance of strain-engineering in tailoring the electronic and vibrational properties but also provide a thorough understanding of the lattice dynamics and mechanical strength of ML-ZnO.  
\end{abstract}


\date{\today}
\maketitle

\section{Introduction}
Since the discovery of graphene \cite{novoselov2004electric} and its unique physical and chemical properties, a lot of interest is drawn toward atomically thin crystals \cite{huang2011graphene, srivastava2012functionalized, yoo2011ultrathin, lin2009operation, novoselov2005two}. Within the continuously growing family of layered materials, those with a sizable band gap have attracted tremendous attention due to their electronic and optical applications potential \cite{wilson1969transition, benameur2011visibility, radisavljevic2011single, pradhan2013intrinsic} and numerous other applications in materials science \cite{yazyev2015mos2, radisavljevic2011integrated, wang2012electronics, johari2011tunable}.     
    
Recent advances in material engineering have shown that many of the binary compounds which crystallize in wurtzite structures can be exfoliated into graphene-like planar monolayer sheets from their bulk counterparts \cite{novoselov2005two, mak2010atomically, geim2013van, rai2018band}. Bulk ZnO which crystallizes in wurtzite structure can be exfoliated to be realized in $2$D-graphene like stable monolayer \cite{claeyssens2005growth}. The single-layer counterpart of ZnO offers attractive physical and chemical properties, for example, wide electronic band gap, high structural stability and spin transport on doping. In experiments, a single-atom-thick ZnO layer was grown on graphene substrates using the atomic layer deposition method \cite{hong2017atomic}. The growth of a monolayer (ML) and few-layers of ZnO on Au (111) have shown that the band gap decreases with the increasing number of layers, from $4.48$ eV for a monolayer to $3.42$ eV for four layers \cite{lee2016tunable}. The existence of a wide band gap assures the potential of ML-ZnO for future applications in various optoelectronic devices. The tunability of the electronic, optical and mechanical properties of $2$D nanomaterials are highly desirable for their potential application in low-power flexible transistors and numerous other opto- and electro-mechanical systems \cite{chu2009strain}. Out of the possible ways to tune the electronic and optical properties of ML-ZnO, strain engineering is an efficient way due to its comparatively easy implementation and reversible mode of application. Our recent study on single-layer MoS$_2$ \cite{chaudhuri2022strain} suggests that the electronic and vibrational properties can be altered effectively and efficiently with mechanical strain. So far, the mechanical strain has been applied successfully in graphene \cite{huang2010probing} as well as other $2$D-materials and the optical, electronic and vibrational properties are tuned efficiently \cite{kumar2013mechanical}.  

In addition, a proper investigation of the phonon band dispersions as a function of strain can give information about the presence or absence of ripples in the nano-sheet and the critical strain before which the structure becomes unstable. Ripples are found to exist not only in free-standing graphene \cite{booth2008macroscopic} but also in graphene grown on various substrates \cite{stoberl2008morphology}. Using the shell elasticity theory, the formation of ripples can be analysed, particularly by examining the behaviour of the out-of-plane acoustic mode or the ZA mode. The dispersion relation of the acoustic modes can be fitted with the model equation $\omega^2 = \text{A}k^4 + \text{B}k^2$, where A and B are related to the bending modulus of the sheet and the applied strain, respectively. Mechanical strains are almost invariably generated within the nano-sheets during their integration on to any substrates. Thus, a proper understanding of the behaviour of ZA mode and hence, the formation or non-formation of ripples with strain application is essential. From the fitting of dispersion of the ZA mode with the model equation, the coefficients of the linear (B) and the quadratic (A) terms can be extracted. Increasing linearity (increase in B) of the ZA mode, which shows dominant quadratic behaviour in the unstrained condition, is used as an indication to conclude that monolayers of graphene \cite{rakshit2010absence} and MoS$_2$ \cite{soni2015ab} may not show rippling in the strained condition.   

Here, in this work, first the choice of the exchange-correlation (XC) functional is made after critically evaluating the geometrical and electronic properties of ML-ZnO with different XC functional and Hubbard parameter (U) in DFT+U calculations. The lattice parameter, electronic properties and the relative position of the Zn-d states obtained from experiments and HSE06 hybrid functional calculations are considered as a reference behind the determination of the appropriate XC functional. We then considered both tensile and compressive strains applied within the plane of graphitic ML-ZnO along different directions, such as armchair (AC), zigzag (ZZ) and homogeneous biaxial strain (BI) to investigate the strain-induced modifications in the electronic and lattice dynamical properties. Due to the high probability of strain development within the monolayer sheet during its growth or integration on any substrate, a thorough and accurate study of the impact of strain on various physical and chemical properties of ML-ZnO is necessary. Furthermore, a qualitative analysis of the possibility of ripple formation with strain and the maximum amount of strain that a single-layer of ZnO can withstand is essential, which lacks in the available literature. In addition, the potential of ML-ZnO to be a good thermoelectric material is identified based on the chemical intuition, and caculated electronic and phonon band structures. The efficient tunability of the electronic and phononic properties with strain application also suggests the use of ML-ZnO as a flexible thermoelectric material, where the thermoelectric efficiency and transport properties can be tuned with strain. Our results not only show the tuning of electronic and phononic properties with a wide range of strain profiles but also provides a detailed understanding of the lattice dynamics and stability limit of ML-ZnO.

\section{Computational details}
First-principles calculations using ab-initio density functional theory (DFT) have been performed as implemented within the Vienna Ab Initio Package (VASP) \cite{kresse1996efficient, kresse1996efficiency} together with projector augmented wave (PAW) potentials to account for the electron-ion interactions \cite{kresse1999ultrasoft}. For the electronic exchange and correlation, LDA+U method is used with a Hubbard-U of $8$ eV for the Zn–$d$ states \cite{kohn1965self}. The choice on the exchange-correlation (XC) functional is made based on a series of calculations on the geometrical and electronic properties of ML-ZnO using DFT+U calculations with different U values and hybrid functional calculations with HSE06 functional. An appropriate U value is necessary for correcting the self-interaction error, and has been used in earlier studies \cite{rakshit2011stability, das2014microscopic}. In order to minimize the interaction between the adjacent layers along the out-of-plane direction, a supercell containing a sufficient vacuum of $20\,\text{\AA}$  is used in all calculations. The internal position of the atoms in the unstrained as well as in the strained cases are relaxed using a conjugate gradient scheme until the forces on the atoms are less than $0.01\, \text{eV/\AA}$. In the hexagonal geometry, a well-converged Monkhorst-Pack \cite{monkhorst1976special} k-point set of $21 \times 21 \times 1$ together with a plane wave cut-off energy of $450$ eV are used for the geometry relaxation, and a much denser converged k-point set is used for the post-relaxation calculations. For the orthorhombic geometry, the k-mesh is correspondingly scaled. The phonon dispersion curves are computed using the finite difference method as implemented within the Phonopy package \cite{togo2015first} with a fixed displacement amplitude of $0.015$ \AA. In order to get the interatomic force constants (IFC) with sufficient accuracy, a large supercell of dimension $4 \times 4 \times 1$ with respect to the primitive unit cell together with a strict energy convergence criterion of $10^{-8}$ eV is used.

\section{Results and Discussion}
\subsection{Unstrained monolayer ZnO}
We first investigate the dependence of the geometrical and electronic properties of monolayer (ML) ZnO on the choice of the XC functional. In order to explore the dependency, we have performed first-principles DFT+U calculations with different U values and hybrid functional calculations on ML-ZnO. Finally, the choice of the appropriate XC functional is made for which the lattice parameter, electronic band gap and the relative position of the Zn-$d$ states agree well with experimental observations. Explicitly calculating the Hubbard-U parameters using the ACBN0 method, an earlier report suggested that a U value as high as $\sim$14 eV for the Zn-$d$ states is required to correct the over-delocalization issue of normal LDA or GGA based DFT calculations \cite{agapito2015reformulation}. In our study, the choice of the XC functional and the U value are made after critically evaluating the electronic and the geometrical properties of ML-ZnO using different XC functional and U values. A summary of the calculated lattice parameter and the band gap with different XC functional and U values together with the experimentally observed ones are presented in Table \ref{tab:Table 0}.

\begin{table}[h]
	\caption{\label{tab:Table 0} The calculated geometrical properties, such as the lattice parameter (a), the Zn-O bond length (d) and the electronic band gap (E$_\text{g}$) of ML-ZnO using different XC functional and U values. The calculated values are compared with the experimentally found values in ref. \cite{lizandara2010molecular}.}
	\begin{tabular*}{\textwidth}{ c @{\extracolsep{\fill}} c c c c c c c }
		\hline
		        & \multicolumn{1}{p{2cm}}{\centering exp \\ ref. \cite{lizandara2010molecular}} & \multicolumn{1}{p{2cm}}{\centering LDA+U \\U = 0 eV} & \multicolumn{1}{p{2cm}}{\centering LDA+U \\U = 4 eV} & \multicolumn{1}{p{2cm}}{\centering LDA+U \\U = 8 eV} & \multicolumn{1}{p{2cm}}{\centering GGA+U \\U = 0 eV} & \multicolumn{1}{p{2cm}}{\centering GGA+U \\U = 4 eV} & \multicolumn{1}{p{2cm}}{\centering GGA+U \\U = 8 eV} \\
		\hline 
		a (\AA) & 3.09 & 3.20 & 3.15 & 3.07 & 3.28 & 3.23 & 3.15 \\
		\hline  
		d (\AA) & 1.784 & 1.848 & 1.821 & 1.773 & 1.893 & 1.869 & 1.821 \\
		\hline
		E$_\text{g}$ (eV) & 3.5 & 1.71 & 2.26 & 2.79 & 1.68 & 2.16 & 2.74 \\
		\hline
	\end{tabular*}
\end{table}

Note that, the experimentally found lattice parameter for graphene like planar ZnO is $3.09$ $\text{\AA}$  with a band gap of $3.5$ eV \cite{lizandara2010molecular}. Depending on the substrates used, for example Au or graphene, different values for the lattice parameter and the band gap of ML-ZnO can be found in literature \cite{hong2017atomic, wu2019controlled, quang2015situ}. Similarly, a wide range of values for the lattice parameter and band gap of ML-ZnO prevail in the literature from theoretical calculations using different XC functional and U values \cite{das2014microscopic, monteiro2021mechanical, kaewmaraya2014strain}. In the absence of experiments conducted on free-standing ZnO monolayer, the values calculated with hybrid functionals can be taken as a reliable reference. The band gap value of ML-ZnO is found to be 3.34 eV using HSE06 functional, which is in good agreement with earlier reports using hybrid functional ($3.25$ eV) \cite{das2014microscopic, rakshit2013indirect} and GW approximation ($3.57$ eV) \cite{tu2010first}, as well as with experimental observations \cite{lizandara2010molecular, quang2015situ}.

\begin{figure}[h!]
	\centering
	\includegraphics[scale=0.35]{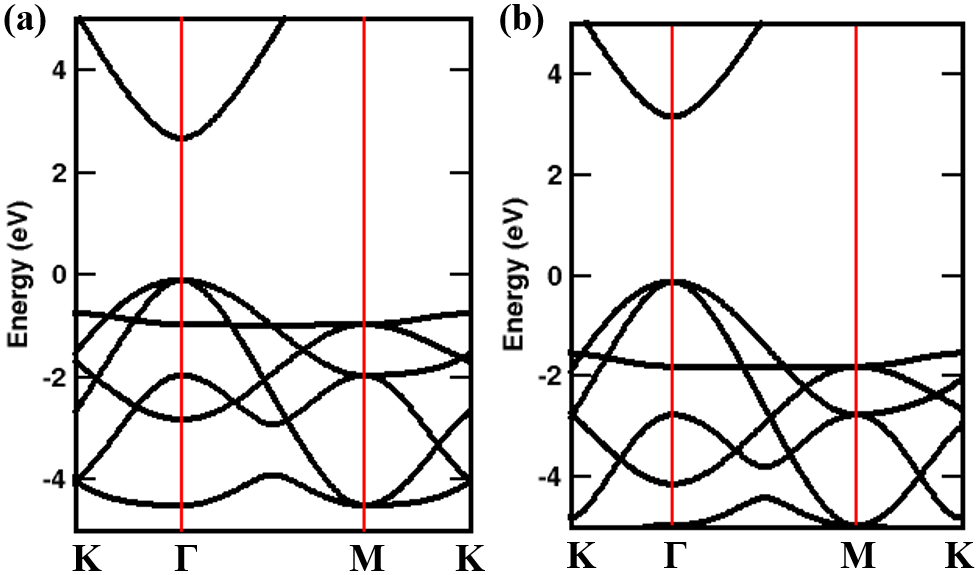}
	\caption{ Electronic band structure of ML-ZnO in the unstrained condition, calculated using (a) LDA+U (U = $8$ eV) and (b) HSE06 functional.} 
	\label{Fig.0}
\end{figure}

It can be seen from Fig. \ref{Fig.0} that the band structures calculated using the LDA+U and HSE06 functional are nearly identical from the qualitative aspect, however, the band gap increases to $3.34$ eV with the HSE06 functional. Using the computationally much less demanding LDA+U (U = $8$ eV) calculations, we obtained a lattice parameter of $3.07 \text{\AA}$ and a band gap of $2.79$ eV for the ZnO monolayer. These results with LDA+U calculations are not only close to the experimentally obtained values, but also agrees well with the hybrid functional calculations. We, therefore, affirm that the choice of the XC functional (LDA) and the value of U ($8$ eV) made here, is legitimate.

\begin{figure}[h!]
	\centering
	\includegraphics[scale=0.4]{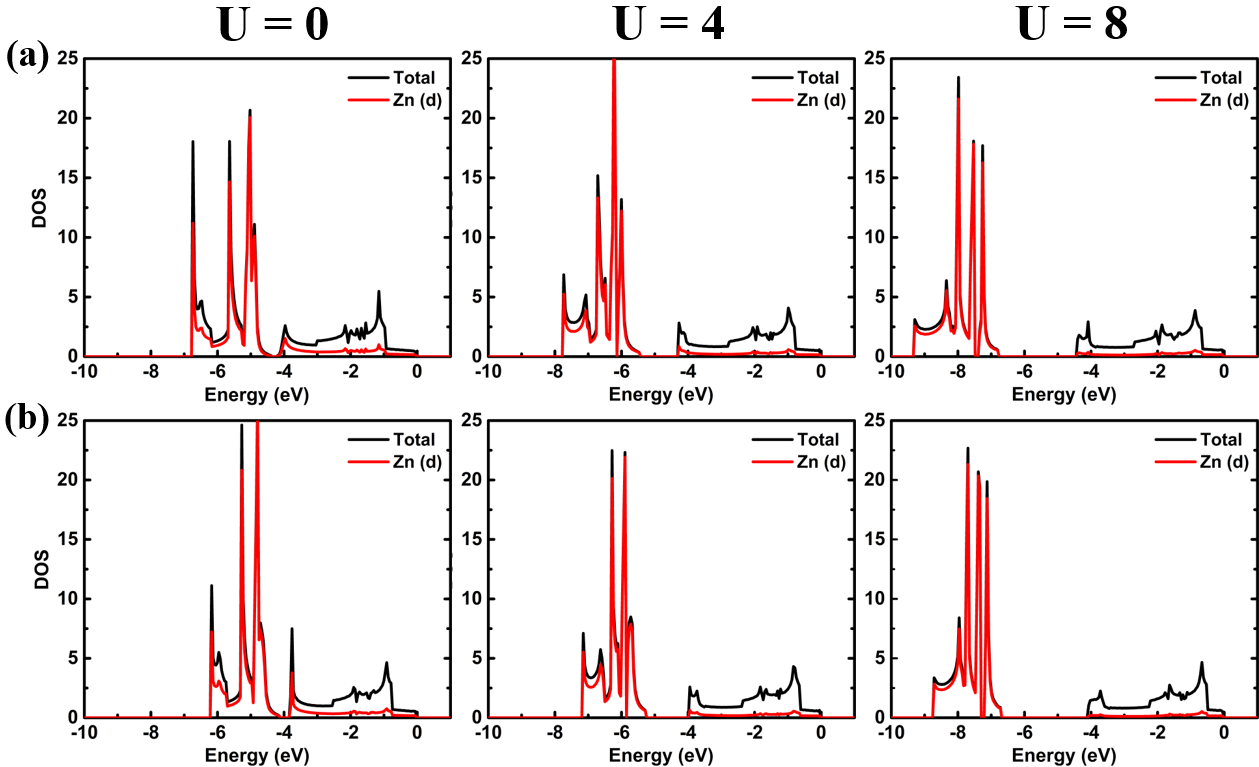}
	\caption{ The electronic density of states (DOS) of ML-ZnO using (a) LDA+U and (b) GGA+U exchange-correlation functional with U = $0$, $4$, $8$ eV. Black and red lines represent the total DOS and the contribution from Zn-$d$ states, respectively. The zero energy (E = $0$ eV) corresponds to the Fermi level.} 
	\label{Fig.01}
\end{figure}

Furthermore, we have used the experimentally determined position of the Zn-$d$ states as a reference behind making the choice of the XC functional and the U value. In the literature, using photoemission experiments on wurtzite ZnO, the position of the Zn-$d$ states is found to be peaked at $7.4$ eV \cite{lim2012angle}. Considering the Zn-$d$ states as semi-core states, we expect the position not to change significantly in graphene like planar structure of ZnO. It has been reported that, hybrid functional calculations, although improve the band gap, lead to a large underestimation of around $2$ eV in the energy value corresponding to the peak of the Zn-$d$ states \cite{rakshit2013indirect}. However, using the LDA+U or GGA+U XC functional, which are computationally much cheaper, a better agreement on the relative position of the Zn-$d$ states is achieved. From the DOS plots shown in Fig. \ref{Fig.01}, it can be seen that with an appropriate value of U, which is $8$ eV in this case, the DOS peaks corresponding to the Zn-$d$ states appear around $7.5$ eV, agreeing well with experiments. With lower U values a large disagreement with experimentally found energy value is observed. This suggests that LDA+U with U = $8$ eV is appropriate for ML-ZnO to correctly describe the geometrical and electronic properties.    

\begin{table}[h]
	\caption{\label{tab:Table 1} Calculated lattice parameters of the hexagonal unit cell and the orthorhombic conventional cell of monolayer ZnO.}
	\begin{tabular*}{0.7\textwidth}{ c @{\extracolsep{\fill}} c c c }
		\hline
		Crystal structure & a (\AA) & b (\AA) & $\gamma$ (degree) \\
		\hline 
		Hexagonal & 3.07 & 3.07 & 120 \\
		\hline  
		Orthorhombic & 3.07 & 5.31 & 90 \\
		\hline    
	\end{tabular*}
\end{table}

\begin{figure}[h!]
	\centering
	\includegraphics[scale=0.5]{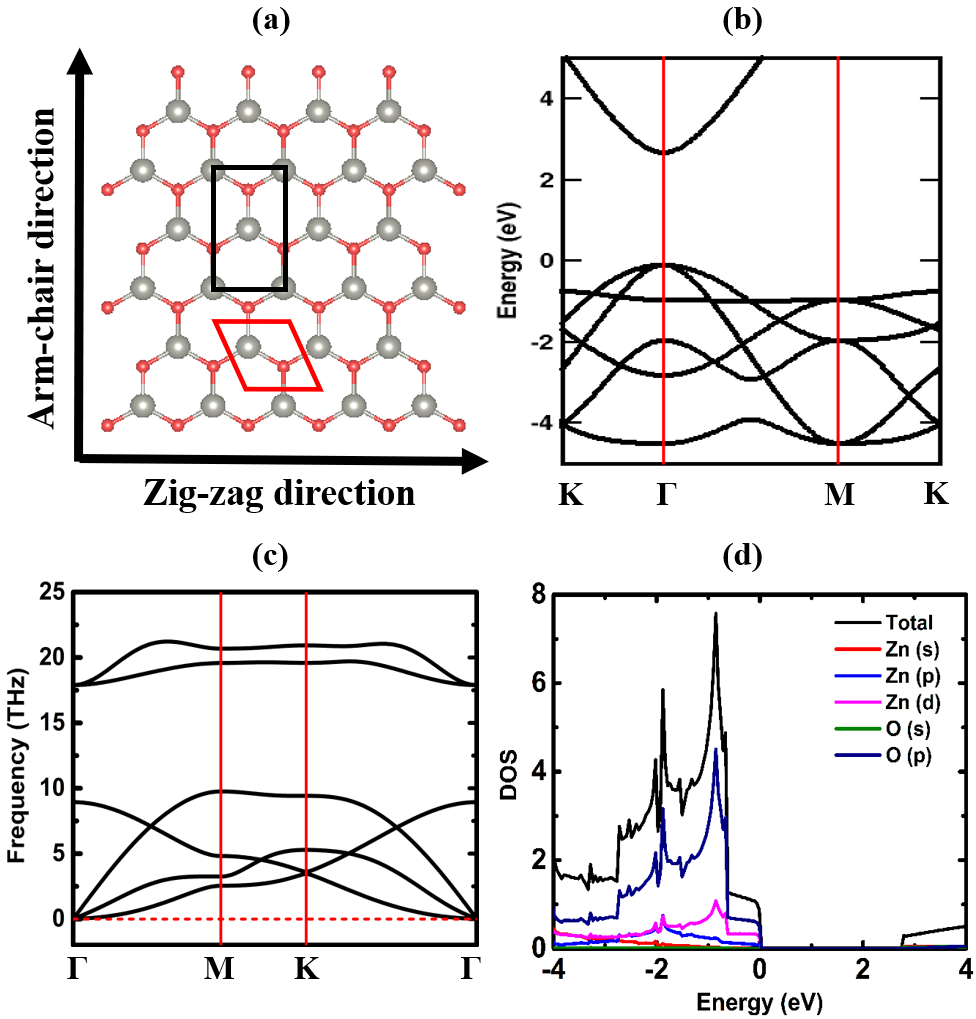}
	\caption{ (a) Crystal structure of monolayer of ZnO (Zn and O atoms are shown in grey and red, respectively). The arrows show the arm-chair and the zig-zag directions. The hexagonal unit cell and the orthorhombic conventional cell are shown in red and black, respectively. (b) Electronic and (c) phonon band structure and (d) density of states (DOS) of unstrained ML-ZnO plotted along the high symmetry path of the irreducible Brillouin zone (BZ).}
	\label{Fig.1}
\end{figure}

Before going into the details of the strain-induced effects on the electronic and phononic properties of ML-ZnO, we first present a brief overview of the calculated geometrical, electronic and lattice dynamical properties of ML-ZnO in the unstrained condition. The relaxed geometry of the ML-ZnO sheet possesses a graphene-like structure with the hexagonal primitive unit cell containing one Zn and one O atom, as shown in Fig. \ref{Fig.1}. In order to apply in-plane tensile and compressive strains independently along the two non-equivalent directions, namely the armchair (AC) and the zigzag (ZZ) directions, an orthorhombic unit cell has been used instead of the hexagonal primitive unit cell. The orthorhombic unit cell contains two Zn and two O atoms, as can be seen in Fig. \ref{Fig.1}. In the chosen orthorhombic unit cell, the AC and the ZZ crystallographic directions are desirably aligned with the orthogonal lattice vectors. The lattice parameters of both the unit cells in the relaxed geometry are given in Table \ref{tab:Table 1}. The lattice parameters calculated herein are found to be in good agreement with earlier reports \cite{das2014microscopic, lalrinkima2019strain}. 

ML-ZnO in the unstrained condition is found to be semiconducting with a band gap of $2.79$ eV with both the valence band maximum (VBM) and the conduction band minimum (CBM) at the same high-symmetry point (at $\Gamma$) of the Brillouin Zone (BZ). Thus resulting in a direct band gap semiconductor in the unstrained condition (see Fig. \ref{Fig.1}). The band gap value agrees well with earlier reports using the same exchange-correlation (XC) functional \cite{das2014microscopic}. From the density of states (DOS), shown in Fig. \ref{Fig.1}(d), it is clear that the valence band edge is mostly composed of O-$p$ orbitals. The phonon band structure of single-layer ZnO in the unstrained condition is also shown in Fig. \ref{Fig.1}. The absence of imaginary phonon branches throughout the BZ confirms the stability of the monolayer ZnO in the graphitic form. The phonon band structure consists of three low energy acoustic branches, namely longitudinal acoustic (LA), transverse acoustic (TA) and out-of-plane acoustic (ZA) and three optical branches (ZO, TO and LO). In the unstrained condition, the LA and TA branches exhibit linear behaviour near the zone centre (at $\Gamma$), whereas the ZA branch shows quadratic behaviour. The band dispersion and the frequencies of the optical branches are in good agreement with previous reports \cite{monteiro2021mechanical}.   

\begin{figure}[h!]
	\centering
	\includegraphics[scale=0.6]{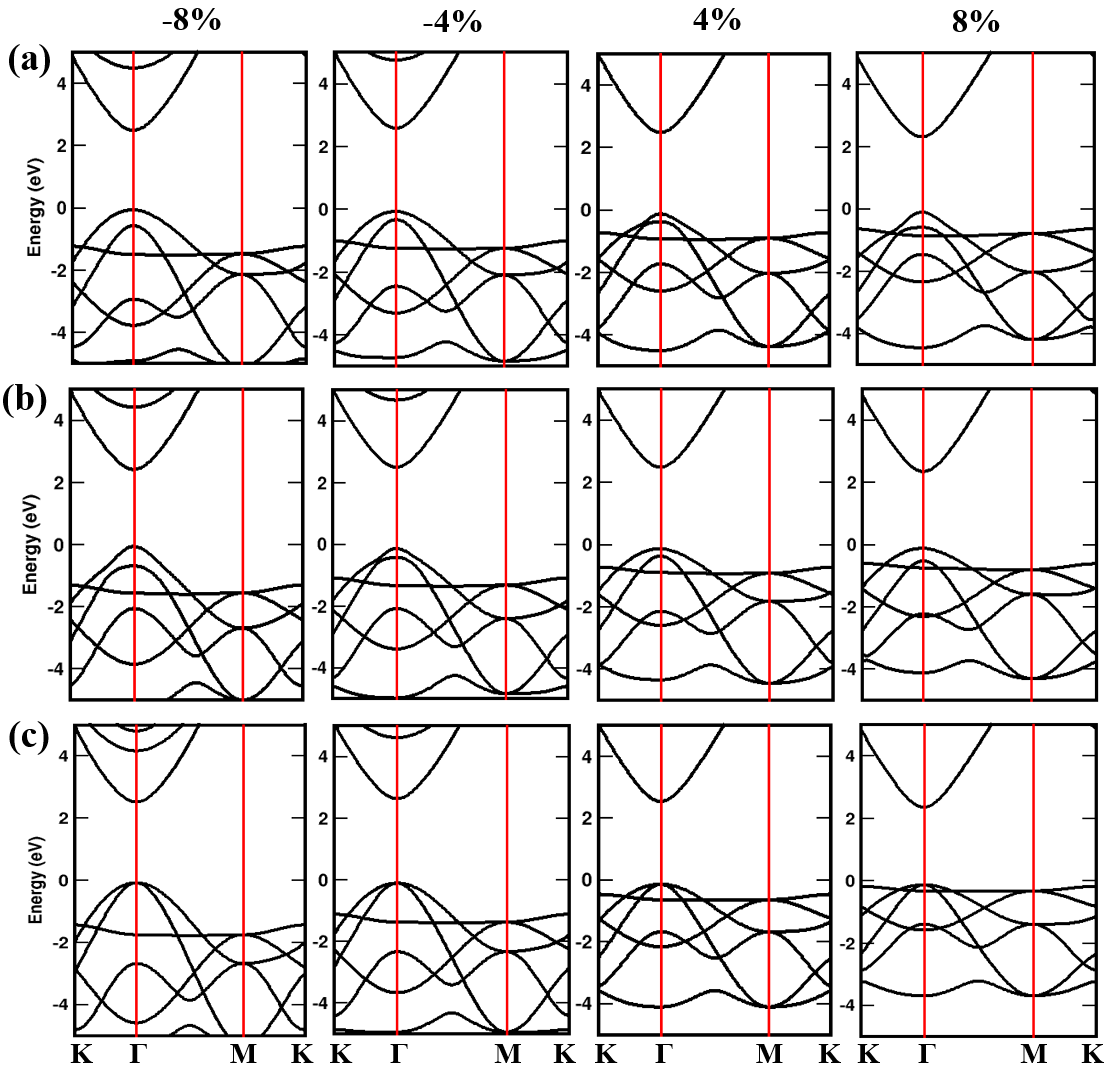}
	\caption{Electronic band structure plots of ML-ZnO at different values of tensile and compressive strains applied along the (a) arm-chair, (b) zig-zag, and (c) homogeneous biaxial strain. The zero energy level (E = $0$ eV) corresponds to the Fermi energy. }
	\label{Fig.2}
\end{figure}

\subsection{Strained monolayer ZnO}
\begin{figure}[h!]
 \centering
 \includegraphics[scale=0.4]{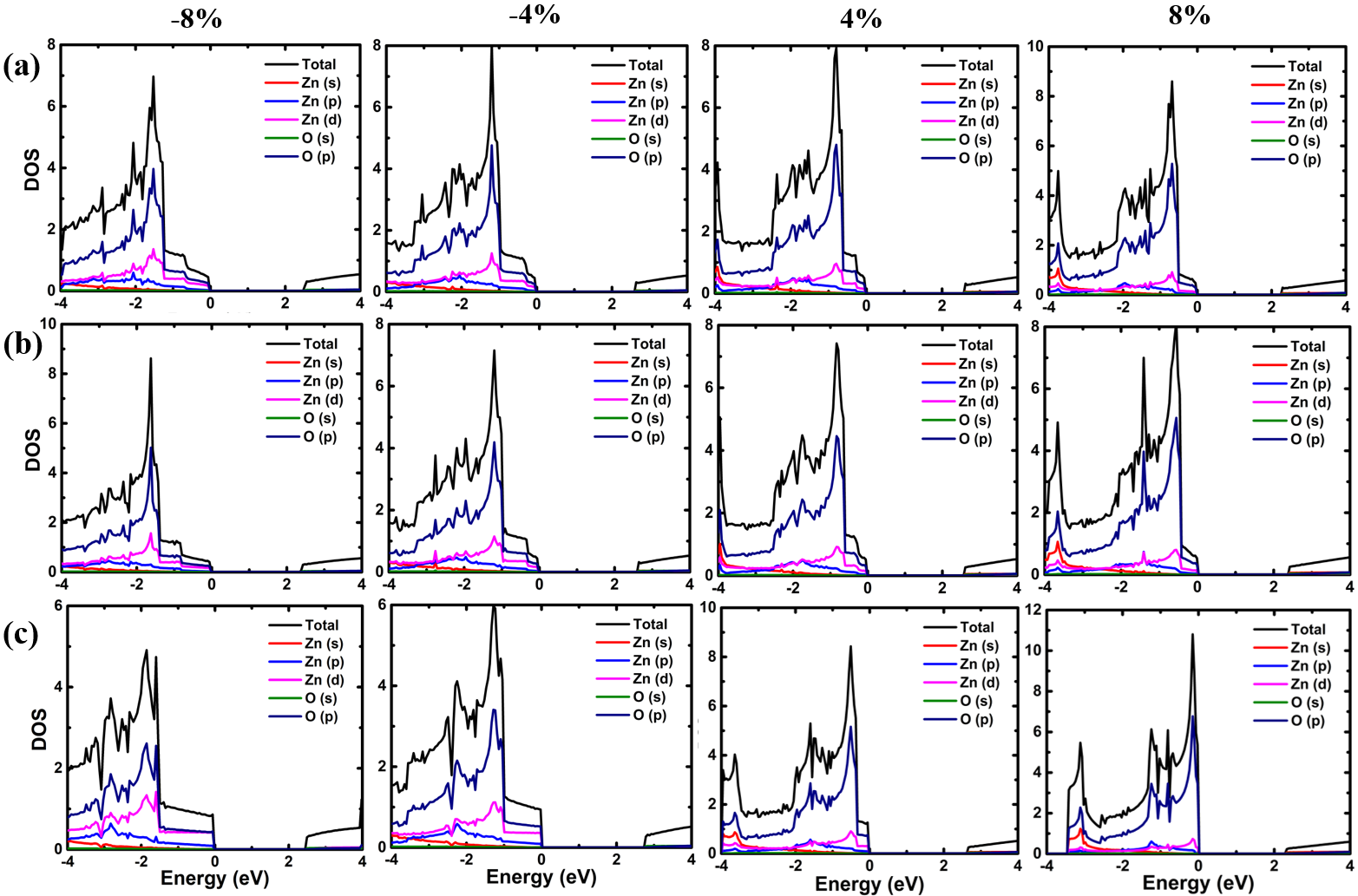}
 \caption{Electronic density of states plots of ML-ZnO at different values of tensile and compressive strains applied along the (a) arm-chair, (b) zig-zag, and (c) homogeneous biaxial strain. The zero energy level (E = $0$ eV) corresponds to the Fermi energy.}
 \label{Fig.3}
\end{figure}

\begin{figure}[h!]
 \centering
 \includegraphics[scale=0.4]{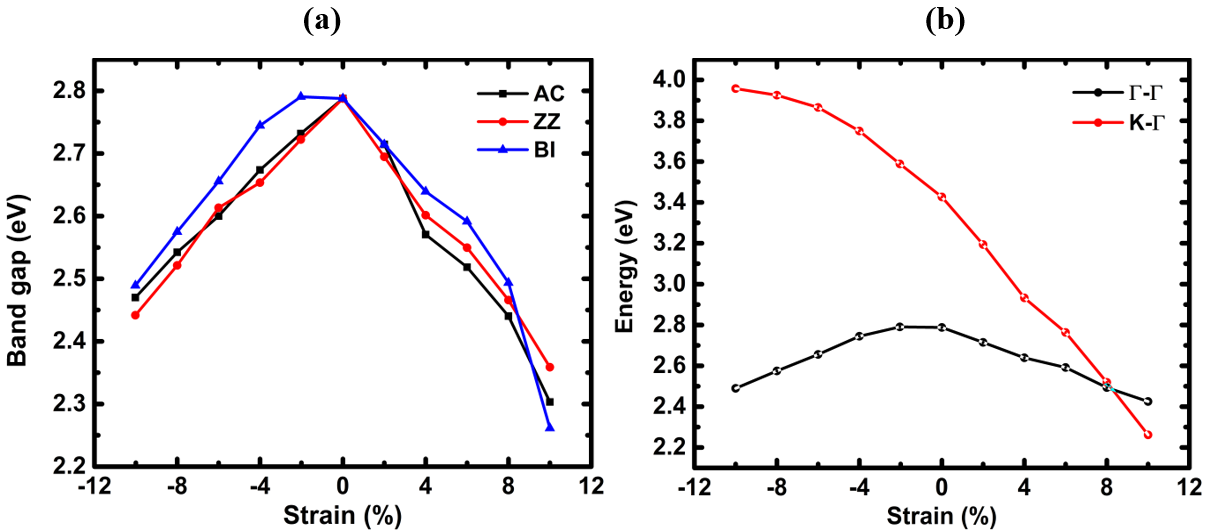}
 \caption{Variation in the (a) band gap, (b) band edge transition energies of ML-ZnO as a function of tensile and compressive strains along different directions.}
 \label{Fig.4}
\end{figure}

To explore the importance of strain-engineering in tuning the electronic properties of ML-ZnO, a series of in-plane tensile and compressive strains have been applied. In-plane tensile and compressive strains are applied along the AC, ZZ, and homogeneous biaxial strain along both the directions. The applied strain is quantified as $\epsilon = \frac{\text{a} - \text{a}_0}{\text{a}_0} \times 100\%$, where a$_0$ and a are the unstrained and strained lattice parameters, respectively. The modifications in the electronic band structure of ML-ZnO under the application of tensile and compressive strains of different magnitude and along different directions are presented in Fig. \ref{Fig.2} and the corresponding DOS of the strained structures are shown in Fig. \ref{Fig.3}. The strain-induced changes in the band gap and the band edge transition energies are summarized in Fig. \ref{Fig.4}. It can be seen that with the application of uniaxial strains along the AC and the ZZ directions, the degeneracy at the top of the valence band (at $\Gamma$) is lifted, whereas the degeneracy prevails with homogeneous biaxial strain. This can be understood from the breaking of hexagonal symmetry under the uniaxial strains, which does not happen under the biaxial strain. Both tensile and compressive strain reduces the band gap of ML-ZnO and the rate of reduction is more or less same for strains along different directions. Although the band gap decreases for both tensile and compressive strains, the mechanism behind the reduction is very different, as has been pointed out in an earlier study \cite{kaewmaraya2014strain}. Within the applied strain range ($\pm 10\%$), the band gap of ML-ZnO reduces from $2.79$ eV to $2.26$ eV for tensile strains and $2.44$ eV for compressive strains. Also, for biaxial strain, a direct ($\Gamma$-$\Gamma$) to indirect (K-$\Gamma$) band gap transition occurs for a strain of $8\%$, as the VBM shifts from the $\Gamma$-point to the K-point. For the application of tensile strains, the valence band edge at K moves upward in energy with no significant change at the conduction band edge. Thereby, the transition of VBM from $\Gamma$ to K occurs (see Fig. \ref{Fig.2} and Fig. \ref{Fig.4}). Due to this upward shift of the valence band edge at K, the density of states (DOS) at the valence band edge is significantly enhanced, as can be seen from Fig. \ref{Fig.3}. Unlike ML-MoS$_2$ or other $2$D materials in general, the rate of reduction of the band gap of ML-ZnO is very less and the direct to indirect band gap transition also occurs at a relatively high value of strain \cite{chaudhuri2022strain, das2014microscopic, chaudhuri2022tensile}. For ML-MoS$_2$, a direct to indirect band gap transition occurs for low values of strain and a semiconductor (band gap $\sim$ $1.8$ eV) to metal transition occurs at a strain of $10\%$ \cite{chaudhuri2022strain}. Thus, it is clear that mechanical strain does not affect the electronic properties of single-layer ZnO the way it impacts other $2$D materials. In order to understand this in greater detail, the electronic structure of ML-ZnO and the orbital contribution of the constituent atoms has been investigated. It is found that the valence band edges at K and $\Gamma$ are composed of the hybridized states of Zn-$d$ and O-$p$ orbitals. Thus, when strain is applied, both the edge states at K and $\Gamma$ would feel the similar shift in energy. Also, the large energy difference between the valence band edge states at $\Gamma$ and K ($\sim$ $0.6$ eV) makes the edge transitions from one k-point to another k-point even more difficult. Therefore, the direct to indirect band gap transition of ML-ZnO occurs at a much higher strain of around $8\%$ compared to the $2\%$ or less in ML-MoS$_2$ \cite{chaudhuri2022strain, chaudhuri2022tensile}.   

Next, the impact of in-plane tensile and compressive strains on the phonon dispersion and lattice dynamics of monolayer ZnO is explored. From an experimental point of view, the application of in-plane compressive strain on a monolayer sheet is a daunting task, whereas in-plane tensile strains can be applied conveniently and reliably. There have been efforts in recent times to apply tensile strain on monolayer or few-layer sheets of various $2$D materials by growing them on a stretchable polymer substrate and then stretching or bending the substrate \cite{pu2013fabrication, rice2013raman, castellanos2013local}. We, therefore, restricted the study of lattice dynamics of ML-ZnO only under tensile strains, keeping the practical aspect in mind. 

\begin{figure}[h!]
	\centering
	\includegraphics[scale=0.5]{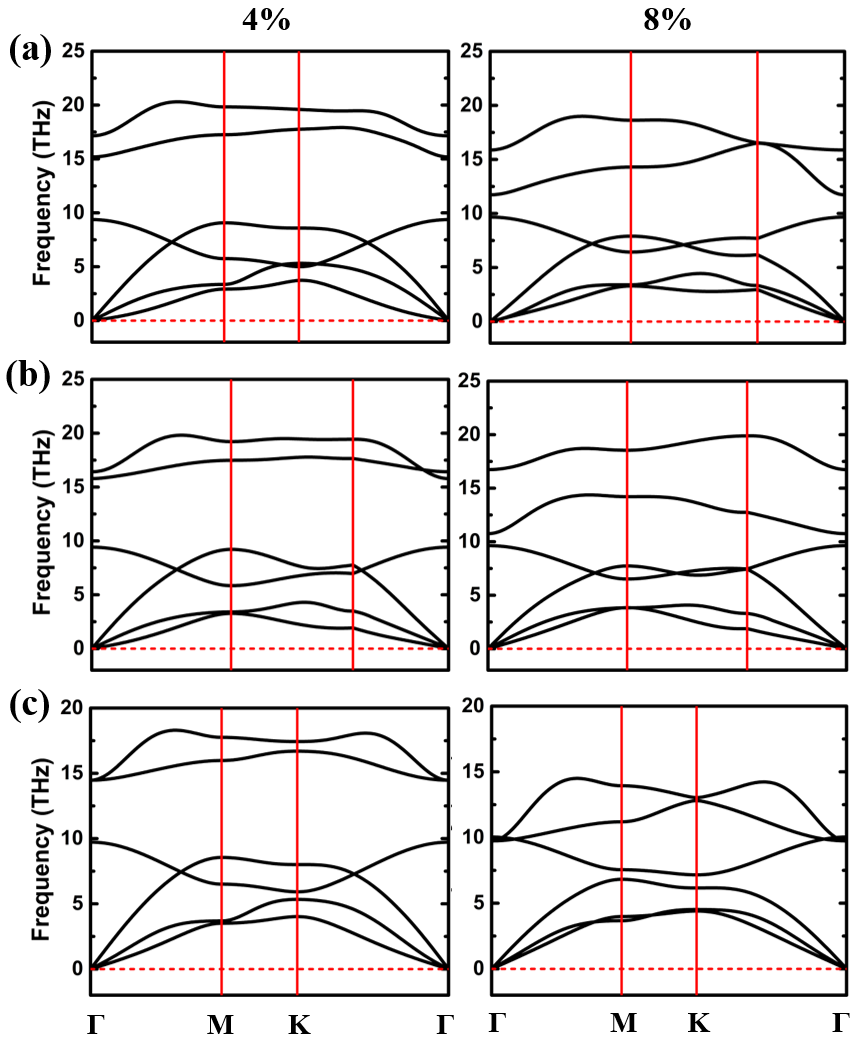}
	\caption{Phonon band structure plots of ML-ZnO at different values of tensile strains ($4\%$ and $8\%$) applied along the (a) arm-chair, (b) zig-zag, and (c) homogeneous biaxial strain.}
	\label{Fig.5}
\end{figure}

To investigate the impact of tensile strains on the phonon properties and lattice dynamics of ML-ZnO and to find out the critical strain at which the structure becomes unstable, in-plane strains along different crystallographic directions (AC, ZZ and BI) have been applied. The modifications in the phonon band structure at the representative strain values are shown in Fig. \ref{Fig.5}. In the unstrained condition, the LO and the TO modes are degenerate at the zone centre ($\Gamma$). With the application of uniaxial tensile strains along the AC and the ZZ direction, the degeneracy is lifted, whereas homogeneous biaxial strain preserves the degeneracy. This is due to the fact that hexagonal crystal symmetry is retained under homogeneous biaxial strain only, and not otherwise. Hence, the LO and TO modes remain degenerate only under biaxial strains. Unlike ML-MoS$_2$, where the frequency of all the optical modes decreases systematically \cite{chaudhuri2022strain} with increasing tensile strain, in ML-ZnO the frequency of the LO and TO modes decreases, whereas the frequency of the ZO mode increases. Such different behaviour of the optical modes i.e. softening of LO and TO modes and hardening of ZO mode with applied strain is due to the different values of the Gruneisen parameter associated with the phonon modes. In order to have a complete understanding of this phenomenon, a thorough study of the mode-resolved phonon parameters is required. The variation in the frequency values of the optical modes with strains of different magnitude and along different directions are summarized in Table \ref{tab:Table 2}.  Another aspect to note here is that the large frequency gap between the ZO mode and the LO and TO modes present in the unstrained condition reduces with the applied tensile strain. For large biaxial strain ($\sim 8\%$), the frequency gap collapses and the ZO mode merges with the LO and TO modes.  

\begin{table}
	\caption{\label{tab:Table 2} Variation in the phonon frequencies (THz) of the optical modes (ZO, LO and TO) at $\Gamma$ for the unstrained as well as $4\%$ and $8\%$ strained structures along armchair (AC), zigzag (ZZ) and biaxial (BI) direction.}
	
	\begin{tabular*}{1.0\textwidth}{ c @{\extracolsep{\fill}} c c c c c c c c c c}
		\hline
		Strain & ZO-AC & LO-AC & TO-AC & ZO-ZZ & LO-ZZ & TO-ZZ & ZO-BI & LO-BI & TO-BI \\
		\hline 
		0\% & 8.96 & 17.89 & 17.89 & 8.96 & 17.89 & 17.89 & 8.96 & 17.89 & 17.89 \\
		\hline  
		4\% & 9.38 & 17.18 & 15.20 & 9.42 & 16.44 & 15.78 & 9.68 & 14.48 & 14.48 \\
		\hline  
		8\% & 9.69 & 15.90 & 11.71 & 9.62 & 16.75 & 10.74 & 10.03 & 9.77 & 9.77 \\
		\hline 
	\end{tabular*}
\end{table} 

The out-of-plane acoustic mode or the flexural mode (ZA) is a characteristic vibrational mode in $2$D materials such as graphene, boron nitride (BN) and transition metal di-chalcogenide (TMDC) sheets. The acoustic phonon modes usually have a linear relationship between the phonon frequency ($\omega$) and the wave vector ($k$). However, from the phonon dispersion curves of ML-ZnO, it can be seen that two of the acoustic modes, namely LA and TA show linear behaviour, while the ZA mode has a significant quadratic ($k^2$) contribution. This mode is generally associated with the rippling of $2$D materials \cite{rakshit2010absence, soni2015ab}. Now, strain is almost invariably generated within the monolayer nanosheet while growing on a substrate or transferring from one substrate to another. Thus, an accurate study of the behaviour of ZA mode with strain application is required to understand the formation or non-formation of ripples within the ML-ZnO sheet. The control of ripples is important to grow ultra-flat sheets of ZnO monolayer on various substrates. The behaviour of the ZA mode can be explained within the shell elasticity theory. The dispersion relation of the acoustic modes is usually given by the model equation $\omega^2 = \text{A}k^4 + \text{B}k^2$, where A and B are related to the bending modulus of the sheet and the amount of strain applied, respectively. To quantify the amount of dependence of $\omega^2$ on the $k^4$ and $k^2$ terms, we fitted the dispersion relation of the ZA mode near the zone centre ($\Gamma$) with the model equation. The values of A and B are extracted for the unstrained as well as strained cases and are presented in Table \ref{tab:Table 3}. It can be seen that values of A decrease, while that of B increase for strains applied along all three directions. This is indicative of the strain-induced hardening of the ZA phonon branch. Due to this continuous increment in the values of B, the ZA mode near $\Gamma$-point becomes nearly linear. Although the values of A and B change for both uniaxial and biaxial strains in similar fashion, the magnitude of decrement is much higher for homogeneous biaxial strain. The large increase in the linear behaviour of the ZA mode with strain suppresses the possibilities of ripple formation or instability of the ML-ZnO sheet grown on a flat substrate.  

\begin{table}
	\caption{\label{tab:Table 3} Variation in the coefficients A and B of the model equation $\omega^2 = \text{A}k^4 + \text{B}k^2$ as a function of strain along AC, ZZ and BI direction. The values of A and B are extracted from the fitting of the model equation to the dispersion of the ZA mode near $\Gamma$-point.}
	
	\begin{tabular*}{1.0\textwidth}{ c @{\extracolsep{\fill}} c c c c c c c}
		\hline
		Strain & A-AC & B-AC & A-ZZ & B-ZZ & A-BI & B-BI \\
		\hline 
		0\% & 8144 & 7.5 & 8144 & 7.5 & 8144 & 7.5 \\
		\hline  
		4\% & 7055 & 131 & 7534 & 217 & 6250 & 336 \\
		\hline  
		8\% & 5127 & 287 & 6426 & 490 & 3544 & 688 \\
		\hline 
	\end{tabular*}
\end{table}

\begin{figure}[h!]
 \centering
 \includegraphics[scale=0.45]{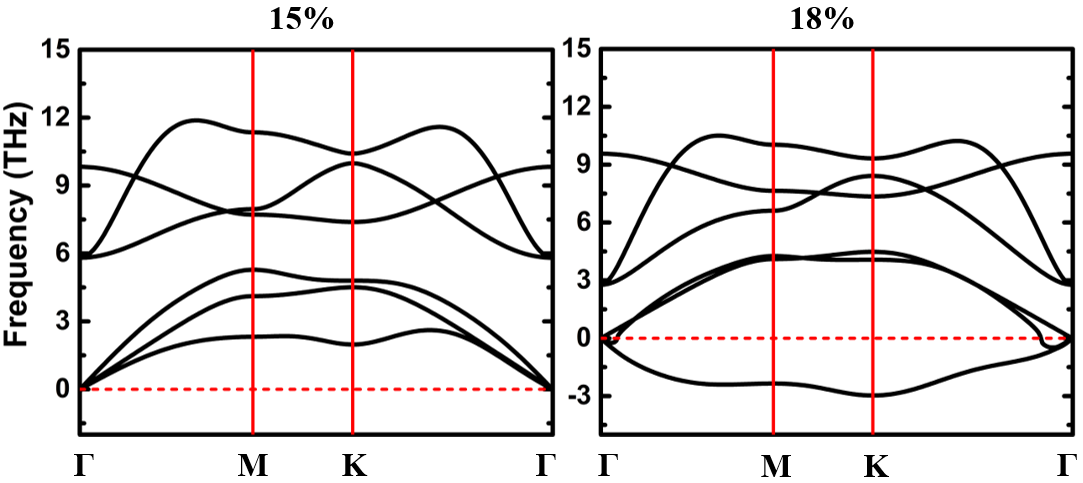}
 \caption{Phonon band structure plots of ML-ZnO under large biaxial strains of $15\%$ and $18\%$. In the $18\%$ strained case the appearence of imaginary phonon branch (in-plane TA mode) throughout the Brillouin Zone can be seen.}
 \label{Fig.6}
\end{figure}

We then went on to examine the behaviour of the ML-ZnO sheet under large biaxial strains. The phonon dispersion curves can be used to find a possible instability in the lattice. The general condition for a stable structure in terms of phonon dispersion is that all phonon modes have real frequency, and imaginary frequency values of any phonon mode within the BZ indicate instability of the structure or a possible phase transition. The phonon band structures of ML-ZnO at large biaxial strain values are presented in Fig. \ref{Fig.6}. Since the impact of the biaxial strain on the lattice dynamics of ML-ZnO is more pronounced compared to the uniaxial strains, it is assumed that the uniaxially strained structures will remain stable at least within the stability limit under biaxial strains. It can be seen that up to a biaxial strain of $15\%$ no imaginary phonon branches emerge within the BZ. Starting from $18\%$ biaxial strain, an acoustic phonon mode, namely the in-plane TA mode, becomes imaginary throughout the BZ. Thus, it can be assumed that the structure of ML-ZnO will become unstable if a biaxial tensile strain of $18\%$ or higher is applied. For graphene also, the emergence of an imaginary phonon mode is observed due to the softening of an acoustic mode at a biaxial strain of $15\%$ \cite{rakshit2010absence}. 

Finally, the promising signs of ML-ZnO as a good thermoelectric material is identified in an intuitive way based on the calculated electronic and phononic properties. The bonding character of ZnO in the wurtzite structure is primarily covalent with a significant contribution from ionic bonding. This results in a large direct band gap in the unstrained condition of ML-ZnO. The semiconducting nature of single-layer ZnO makes it applicable for thermoelectric applications in the first place. With strain application, the band gap reduces at a slow rate and ML-ZnO remains a semiconductor throughout the strain range unlike ML-MoS$_2$, where a semiconductor to metal transition is seen \cite{chaudhuri2022strain, chaudhuri2022tensile}. With tensile strain the bondlength between Zn and O increases, thereby the orbital overlap between the Zn-$d$ and O-$p$ orbitals decreases. Therefore the dispersion of the valence band top decreases, resulting in the increased density of states (DOS) at the valence band edge, as can be seen from Fig. \ref{Fig.3}. Enhanced DOS at the band edges usually result in an increased Seebeck coefficient and thus improve the overall thermoelectric performance. Now, due to the low-dimensional structure and the resulting interface phonon scattering and phonon confinement effects, ML-ZnO will offer low lattice thermal conductivity ($\kappa_L$) compared to its bulk counterpart or any other bulk structures, in general. Low $\kappa_L$ is quintessential for efficient thermoelectric performance. Furthermore, the large frequency gap between the acoustic and optical modes in the unstrained condition ceases to exist with increasing tensile strain and the modes nearly merges at $8\%$ biaxial tensile strain. Therefore, the acoustic-optical phonon scattering will increase with tensile strain and the $\kappa_L$ will be further lowered. Thus, based on the analysis of the electronic and phononic properties of ML-ZnO in the unstrained as well as strained conditions, it is clear that the signs of ML-ZnO as a potential thermoelectric material are encouraging. In some recent studies \cite{ul2021thermoelectric, zhao2018study}, the prospect of ML-ZnO as a thermoelectric material has been explored in the pristine condition based on first-principles calculations. Further studies on the thermoelectric and transport properties of ML-ZnO in the unstrained as well as strained condition will reveal its actual potential.

\section{Conclusion}
In summary, first-principles DFT calculations have been performed to investigate the impact of strain on the electronic and lattice dynamical properties of ML-ZnO. It is found that mechanical strain can effectively modify both the electronic and phononic band structures. With the application of both tensile and compressive strains along different directions (AC, ZZ and BI), the band gap of ML-ZnO decreases. For large biaxial tensile strain, a direct to indirect band gap transition occurs due to the shifting of the VBM from $\Gamma$ to K. The slow rate of decrease of band gap and large required strain for direct to indirect band gap transition, compared to other $2$D materials, are analysed through the investigation of orbital contributions to the electronic structure. With strain, two of the three optical phonon modes soften, while the ZO mode hardens. The ZA mode which shows quadratic behaviour with the phonon wave vector in the unstrained condition turns linear with increasing strain. The behaviour of the ZA phonon mode with strain have been analysed using shell elasticity theory to understand the role of strain in the ripple formation in a monolayer ZnO sheet. The increasing linearity of the ZA mode with strain indicates that single-layer ZnO can be grown on a flat substrate without ripple or resulting instability. Furthermore, the behaviour of the ML-ZnO sheet under large biaxial strain is tested to find out the critical strain value at which the structure becomes unstable. At a homogeneous biaxial strain of $18\%$, the monolayer sheet becomes unstable with the emergence of an imaginary phonon mode. Finally, the clear indications of ML-ZnO to be a good thermoelectric material and the improvement in thermoelectric performance with strain application is predicted based on the analysis of the calculated electronic and phononic properties. Our results may pave a very effective way for the tailoring of electronic and vibrational properties of ML-ZnO and thus enabling the device tunability. 

\section*{Conflict of Interest}
The authors declare that they have no conflict of interest.

\begin{acknowledgments}
	The first-principles calculations are performed using the supercomputing facility of IIT Kharagpur established under National Supercomputing Mission (NSM), Government of India and supported by the Centre for Development of Advanced Computing (CDAC), Pune. SC acknowledges MHRD, India, for financial support.
\end{acknowledgments}

\bibliography{biblio}

\end{document}